# Coupling between Quantum Hall Edge Channels on Opposite Sides of a Hall Bar

Ngoc Han Tu[a], Masayuki Hashisaka[a,b], Takeshi Ota[a], Yoshiaki Sekine[a], Koji Muraki[a], Toshimasa Fujisawa[b], and Norio Kumada[a,*]

[a]NTT Basic Research Laboratories, NTT Corporation, 3-1 Morinosato-Wakamiya, Atsugi, 243-0198, Japan

[b]Department of Physics, Tokyo Institute of Technology, 2-12-1 Ookayama, Meguro, Tokyo 152-8551, Japan

Corresponding author at: NTT Basic Research Laboratories, NTT Corporation, 3-1 Morinosato-Wakamiya, Atsugi, 243-0198, Japan
E-mail address: kumada.norio@lab.ntt.co.jp

We investigate the coupling between quantum Hall (QH) edge channels (ECs) located at opposite sides of a 50-μm-wide Hall bar by exciting a charged wavepacket in one EC and detecting time-dependent current in the other EC. In a QH state, the current shows a peak followed by a dip, demonstrating the existence of capacitive coupling across the incompressible two-dimensional electron system (2DES). The observed magnetic field dependence of the amplitude and time delay of the current suggests that the capacitance is affected by the presence of localized states. We also show that the dominant manner of the coupling changes gradually as the system changes between the QH and non-QH states.

## 1. Introduction

In a quantum Hall (QH) state, electrons immersed in a perpendicular magnetic field propagate in chiral channels along the periphery of the two-dimensional electron system (2DES). The chiral edge channels (ECs) are considered ideal one-dimensional (1D) channels and used to perform quantum transport experiments. Examples include electron interference experiments based on concepts derived from quantum optics [1-5]. An important difference from optics is the presence of the Coulomb interaction. It has been shown that interactions between co-propagating ECs located on one side of a sample and separated by a narrow incompressible strip induce decoherence [6, 7] and energy relaxation [8, 9]. ECs are also used to investigate properties of Tomonaga-Luttinger liquid [10-13]. In these experiments, interactions between two ECs close to each other were used to induce spin-charge separation [10, 13] and charge fractionalization [11, 12].

In contrast to these investigations of interaction effects between adjacent ECs, interactions between ECs on opposite sides of a standard Hall bar sample with a typical width of 10 μm are routinely ignored because of larger distance. However, high-frequency transmittance measurements have shown that interactions between ECs separated by ~ 100 μm are not negligible [14]. Indeed, a recent experiment using a system composed of two ECs separated by etching has demonstrated that the strength of the interactions decreases slowly with the distance in logarithmic law, and it decreases only by a factor of three when the distance increases by more than two orders of magnitude from 0.3 to 50 μm [12]. These results imply that the interaction across the bulk 2DES, which has been investigated for a narrow constriction [15, 16], should be taken into account to better understand quantum transport even in a standard Hall bar sample.

In this work, we investigate how the electronic state of the bulk 2DES affects the interaction between ECs propagating on the opposite sides of a sample by time-resolved transport measurements. We used graphene sample with a width of 50 μm. We selectively injected a charged wavepacket into an EC on one side (EC1) and measured the current in ECs on both sides in the time domain through two ohmic contacts Det1 and Det2 connected to EC1 and the other EC (EC2), respectively. The existence of interactions is confirmed by the observation of finite current at Det2. In a QH state, the current at Det2 shows a peak followed by a dip as a function of time, while its time integration is zero. This indicates that the two ECs are capacitively coupled across the incompressible 2DES. The strength of the capacitive coupling varies with magnetic field even in the region with vanishing longitudinal resistance $R_{xx}$. We suggest that localized states inside the incompressible 2DES enlarge the coupling strength. In the non-QH state, on the other hand, the Det2 current becomes a single peak as expected for the conductive coupling through the compressible 2DES. For completeness, we carried out similar measurements using a 2DES formed in GaAs, which is often used in quantum transport experiments. In a GaAs sample, contributions of the capacitive and conductive couplings oscillate following $R_{xx}$. These results indicate that presence of the inter-EC coupling is a common feature of QH systems.

## 2. Material and methods

Most of the data presented were obtained using graphene grown by thermal decomposition of a 6H-SiC(0001) substrate. Figure 1(a) schematically shows the device structure and experimental setup. Graphene was etched to form a 50-μm-wide and 350-μm-long bar region. EC1 and 2 propagate along the top and bottom edges of the bar region, respectively. Six Cr/Au ohmic contacts were deposited on graphene. Two of them located at the downstream of EC1 and 2 from the bar region serve as Det1 and 2, respectively. After the formation of ohmic contacts, the surface of the graphene was covered with 100-nm-thick hydrogen silsesquioxane and 60-nm-thick $SiO_2$ insulating layers. An injection gate having a $10 \times 10$ μm$^2$ overlap with the patterned graphene was deposited on the insulating layer. The carriers are electrons, and the density is about $5 \times 10^{11}$ cm$^{-2}$. The low-temperature mobility is about 12000 cm$^2$/Vs.

We also fabricated a similar device with 20-nm-thick GaAs/Al$_{0.3}$Ga$_{0.7}$As quantum well. A 15-μm-wide and 350-μm-long bar region is formed by wet etching, and ohmic contacts were made by alloying with AuGeNi. The geometry other than the bar region is the same as that of the graphene sample. The electron density is $1.2 \times 10^{11}$ cm$^{-2}$, and the low-temperature mobility is $2.1 \times 10^6$ cm$^2$/Vs.

Measurements were performed at 1.5 K for the graphene sample and about 100 mK for the GaAs sample. $R_{xx}$ across the bar region was measured using a standard low-frequency lock-in technique. Magnetic field $B$ up to 10 T was applied from the back of the sample so that the chirality of ECs is clockwise at $B > 0$ T as shown in Fig. 1(a). A charged wavepacket excited by a voltage step with the rise time of about 100 ps applied to the injection gate propagates as 2D plasmons at $B = 0$ T or edge magnetoplasmons (EMPs) [17-22] at high $B$. From the rise time of the injection pulse, geometry of the injection gate, and the measured plasmon velocity, the width of the plasmon pulse is estimated to be a few hundred micrometers at Landau level filling factor ν = 2 [12] and wider for larger ν. Plasmons reaching Det1 and Det2 are detected as time-dependent current using a sampling oscilloscope through an amplifier with 13.5 GHz bandwidth (model: Picosecond Pulse Labs 5840B). The red thin trace in Fig. 1(b) shows the measured current at Det1, which comprises the plasmon signal and unwanted crosstalk. We set the time origin $t = 0$ at the first peak of the crosstalk because it propagates with the speed of light and its time delay is negligible. The crosstalk component (black trace) can be obtained by measuring the current at inverted magnetic field $B = -10$ T, where the chirality of the ECs is reversed and excited charges do not arrive at Det1 and 2. Then, by subtracting the measured current at $B = -10$ T from those for $B \geq 0$, the crosstalk can be eliminated (red thick trace) [23]. The crosstalk is subtracted for all the current data plotted hereafter.

## 3. Results and discussion

Figures 2(a) and (b) show color-scale plots of the current at Det1 and 2 ($I_{\text{Det1}}$ and $I_{\text{Det2}}$), respectively, as a function of $t$ and $B$. In the $\nu = 2$ QH state for $B \gtrsim 2$ T, where $R_{xx}$ is vanishingly small [Fig. 2(c)], $I_{\text{Det2}}$ shows a peak followed by a dip. We also plotted the integrated current in Fig. 2(d). Net current on Det2 obtained by $\int_0^{10\,\text{ns}} I_{\text{Det2}}\, dt$ is almost zero, consistent with the vanishing $R_{xx}$. As discussed below, this behavior can be explained by a capacitive coupling across the incompressible bulk 2DES. Strength of the capacitive coupling can be evaluated by the area of the current dip obtained by $\frac{1}{2}\int_0^{10\,\text{ns}}(|I_{\text{Det2}}| - I_{\text{Det2}})\, dt$, which becomes a maximum at $B$ close to the boundary to the non-QH state. In the non-QH state for smaller $B$, both $I_{\text{Det1}}$ and $I_{\text{Det2}}$ show a single peak. In this regime, $\int_0^{10\,\text{ns}} I_{\text{Det2}}\, dt$ is finite, indicating that electrons are transmitted to EC2 as expected for the conductive coupling across the compressible bulk with finite $R_{xx}$.

We now focus on the results inside the QH effect regime ($B \gtrsim 2$ T). For epitaxial graphene grown on SiC substrate, charge transfer from the underlying substrate to graphene tends to pin the filling factor to $\nu = 2$, leading to exceptionally wide QH plateau [24, 25]. This allows us to investigate the behavior inside the QH state in detail.

The appearance of a peak in $I_{\text{Det1}}$ and a peak followed by a dip in $I_{\text{Det2}}$ can be explained by the capacitive coupling between counter-propagating ECs in the bar region [11, 12, 26]. An EMP wavepacket with charge $Q$ in EC1 induces charge $-rQ$ in EC2, and this forms a coupled EMP wavepacket in the bar region propagating to the downstream of EC1 [inset of Fig. 3(d)]. Simultaneously, because of the charge conservation, charge $rQ$ is excited in EC2, which propagates to the downstream of EC2 and is detected by Det2 as a current peak. When the coupled EMP wavepacket arrives at the right end of the bar region, charge $-rQ$ in EC2 is reflected with induced charge $r^2Q$ in EC1, while charge $(1-r^2)Q$ is transmitted and appears as a current peak at Det1. The subsequent process at the left end leads to negative current with charge $-r(1-r^2)Q$ at Det2. The value of $r$ ($0 < r < 1$) approaches unity in the strong inter-EC coupling limit. Note that after similar multiple processes, the total charge arriving at Det1 and 2 should eventually become $Q$ and zero, respectively.

In Figs. 3(a) and (b), $I_{\text{Det1}}$ and $I_{\text{Det2}}$ for several values of $B$ are plotted, respectively. At $B \gtrsim 4$ T, where the system is deep in the QH regime, the current profile does not show significant $B$ dependence. For smaller $B$, the amplitude and time position of the current peaks and dip vary with $B$. As $B$ is decreased to the boundary with the non-QH state, the peak in $I_{\text{Det1}}$ becomes smaller, while the peak and dip in $I_{\text{Det2}}$ becomes larger. The increase in the dip amplitude corresponds to the increase in $\frac{1}{2}\int (|I_{\text{Det2}}| - I_{\text{Det2}})dt$ plotted in Fig. 2(d). In the same $B$ range, the time position of the peaks and dip shifts to a larger value with decreasing $B$.

Figure 3(c) shows $r$ as a function of $B$. The values of $r$ are obtained by fitting calculated traces to $I_{\text{Det1}}$ and $I_{\text{Det2}}$ [inset of Figs. 3(a) and (b)]. In the calculation, we used the plasmon dispersion shown in Ref. [12, 27, 28] and reproduced $I_{\text{Det1}}$ and $I_{\text{Det2}}$ using $r$ as a tunable parameter (details of the calculation are given in the supplementary information). Figure 3(c) also includes the velocity $v$ of the coupled EMP mode calculated from the time difference between the peaks in $I_{\text{Det1}}$ and $I_{\text{Det2}}$; since the distance from the bar region to Det1 and Det2 is designed to be the same, the time difference corresponds to the time of flight of the coupled EMP wavepacket in the bar region. As $B$ is decreased below 4 T, $r$ increases and velocity decreases.

We estimate the coupling strength using a chiral distributed-element circuit model composed of two transmission lines with impedance $Z = \sigma_{xy}^{-1}$, channel capacitance $C_{ch}$ mainly arising from carrier interactions, and inter-channel capacitance $C_x$ [inset of Fig. 3(c)]. According to the model [27], $r$ and $v$ are determined by the intra- and inter-EC interactions and can be represented by the capacitances as

$$r = \frac{C_{ch}+C_x-\sqrt{C_{ch}^2+2C_{ch}C_x}}{C_x} \quad (1)$$

and

$$v = \frac{\sigma_{xy}}{\sqrt{C_{ch}^2+2C_{ch}C_x}}. \quad (2)$$

Figure 3(d) shows $C_{ch}$ and $C_x$ calculated based on Eqs. (1) and (2) using the measured $r$ and $v$. $C_x$ increases with deceasing $B$, while $C_{ch}$ is almost unchanged with $B$.

We ascribe the $B$ dependence of $C_x$ to the change in the landscape of localized states inside the incompressible 2DES. The localized states capacitively couple to the ECs and serve as conducting islands between them. EMPs passing close to each localized state cause electrostatic induction in it. This effectively shortens the distance between the ECs and thereby increases $C_x$. As $B$ is decreased to the boundary with the non-QH state, the area of localized states grows and $C_x$ becomes larger. On the other hand, deep in the QH state, the filling factor is pinned at $\nu = 2$ by the charge transfer from the substrate to graphene, and the landscape of localized states should be almost independent of $B$. This explains the constant $C_x$ for $B \gtrsim 4$ T. Note that since the width of the EMP pulse is much larger than the typical length scale of potential fluctuations, averaged $C_x$ for the EMP pulse should be independent of the position of the pulse.

Next, we show results around the boundary region between the QH and non-QH states (Fig. 4). As $B$ is decreased across the boundary ($B \sim 2$ T), the behavior of $I_{Det2}$ changes. For larger $B$, $I_{Det2}$ shows a peak and dip structure induced by a capacitive coupling between EC1 and 2. For smaller $B$, on the other hand, it shows a single peak. In this $B$ range, the peak amplitude of $I_{Det1}$ drops. This indicates that electrons transmit between the ECs through the conductive 2DES. At $B = 1.4$ T, the dip in $I_{Det2}$ still exists, but its amplitude is much smaller than the peak amplitude. This indicates that both the capacitive and conductive couplings contribute to the inter-EC coupling in the transition region.

Finally, we show results for a GaAs sample. Since ECs in GaAs systems are used for quantum transport experiments and plasmonic applications [29], it is valuable to verify the consistency of the behavior in graphene and GaAs. A color-scale plot of $I_{Det2}$ as a function of $t$ and $B$ is plotted in Fig. 5(a). $I_{Det2}$ shows a peak and dip in QH states but only a peak in non-QH states. This indicates the presence of capacitive coupling in QH states and dominant conductive coupling in non-QH states. The contribution of the capacitive and conductive couplings changes with $R_{xx}$ as manifested by the oscillation of the integrated current [Figs. 5(d) and (e)]. These results demonstrate that the observed inter-EC coupling across a 2DES is a common feature of QH systems.

## 4. Conclusions

We investigated the coupling between ECs on opposite sides of a Hall bar device by high-frequency transport measurements. Our results demonstrate that ECs couple capacitively in QH states with vanishing $R_{xx}$. We show that the strength of the capacitance varies with $B$ even inside the QH state. It becomes stronger as the system moves towards the boundary to a non-QH state. We ascribed the enhancement of the capacitance to the expansion of localized states. These results imply that it is essential to include the coupling across a bulk 2DES to better understand charge transport in ECs.

**Acknowledgements**
We thank K. Sasaki for fruitful discussion and A. Tsukada for sample fabrication. This work was supported by JSPS KAKENHI Grant Number JP15H05854.

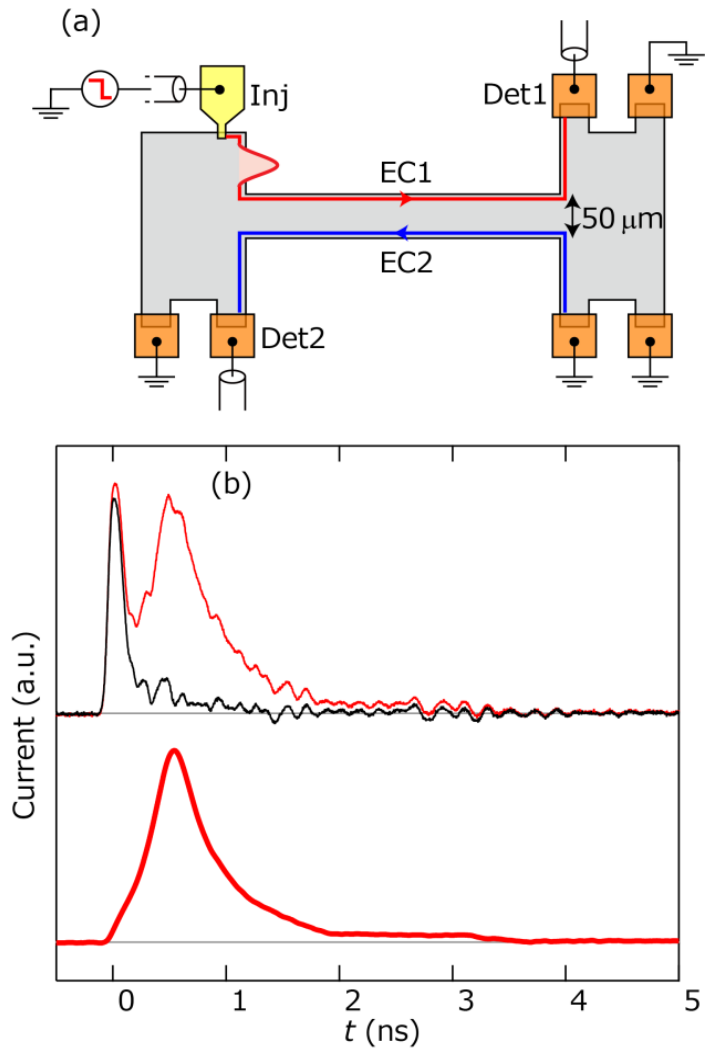

FIG. 1. (a) Schematic of a sample. An injection gate (yellow) and ohmic contacts (orange) are patterned on graphene (grey). High-frequency lines are connected to the injection gate and two ohmic contacts labeled Det1 and Det2. (b) Measured current as a function of $t$ at Det1 at $B = 10$ T (red thin trace) and $-10$ T (black thin trace). By subtracting the current at $B = -10$ T from that at 10 T, crosstalk can be eliminated (red thick trace).

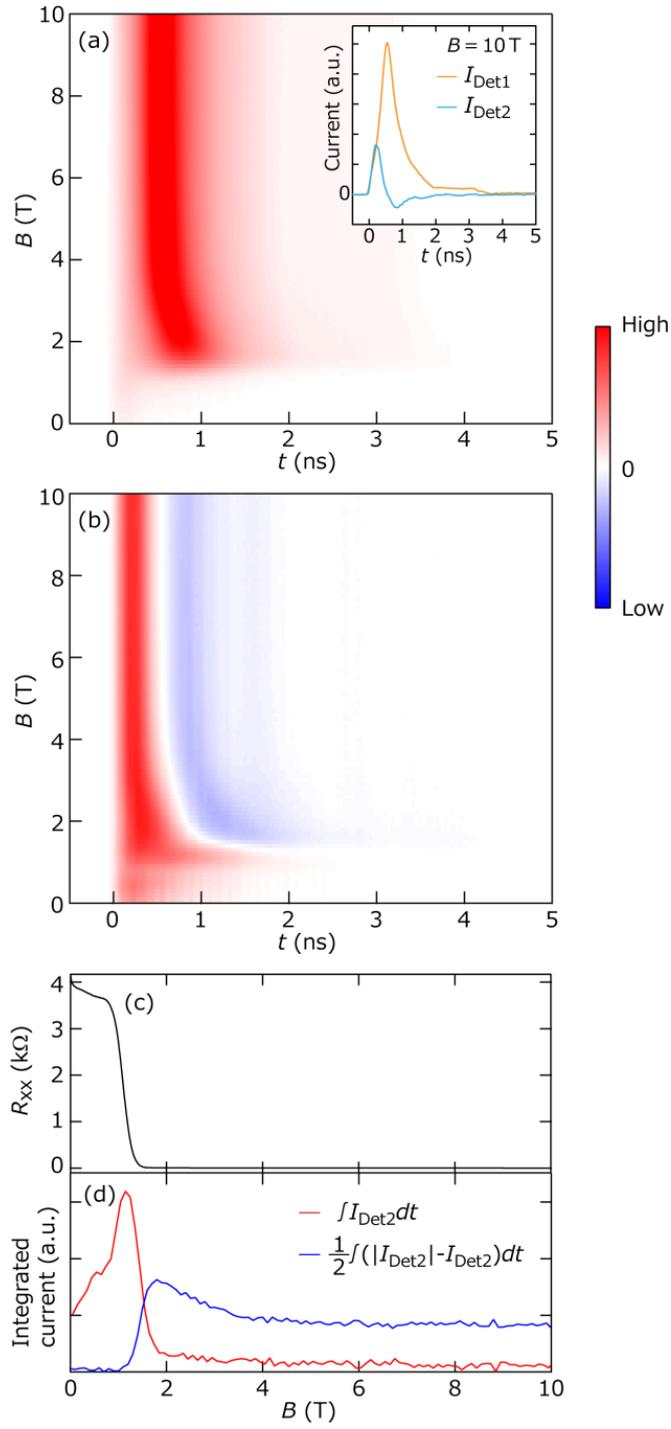

FIG. 2. (a) and (b) $I_{Det1}$ and $I_{Det2}$ as a function of $t$ and $B$, respectively. The color-scale is different between (a) and (b). Inset of (a) shows $I_{Det1}$ and $I_{Det2}$ at $B = 10$ T. (c) and (d) $R_{xx}$ and the integration of $I_{Det2}$ as a function of $B$, respectively.

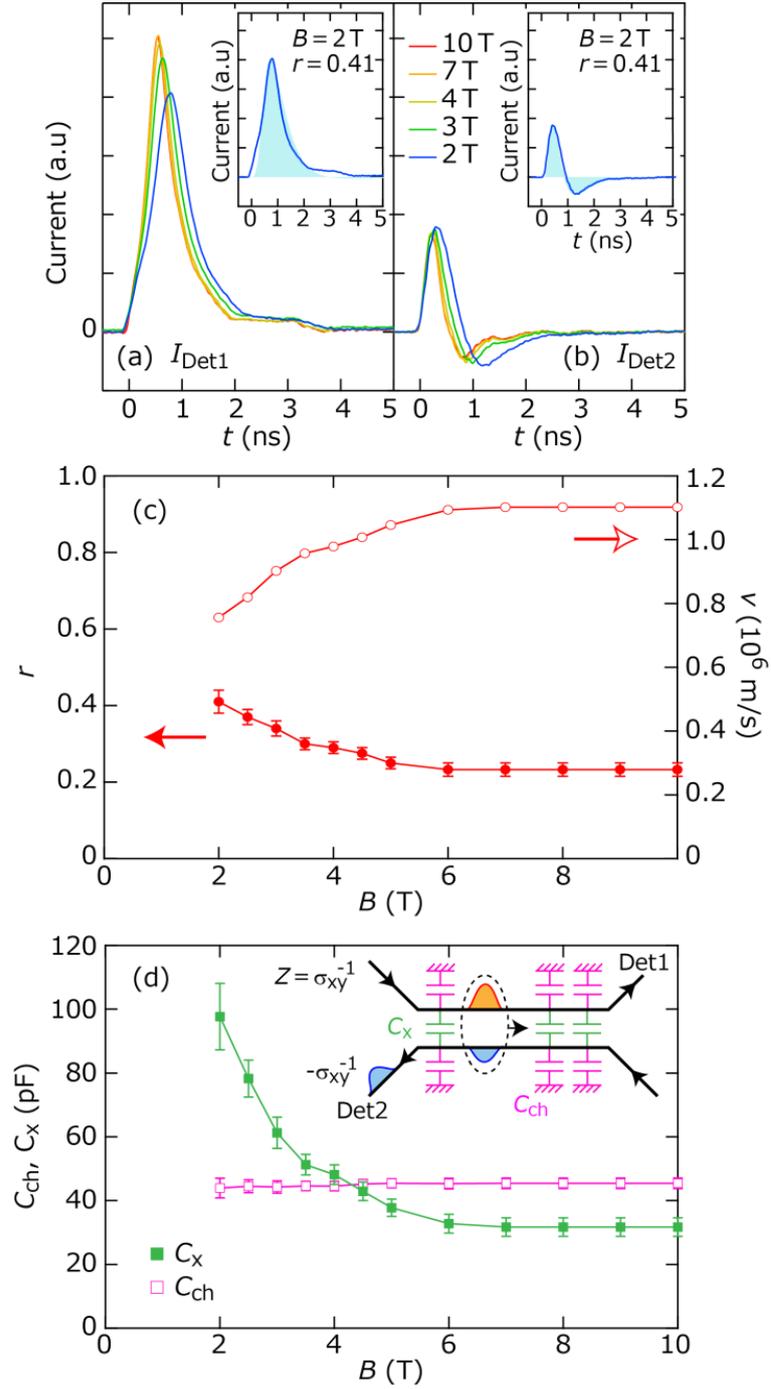

FIG. 3. (a) and (b) $I_{Det1}$ and $I_{Det2}$ as a function of $t$, respectively, for several values of $B$ in the QH state. Insets show the current traces for $B = 2$ T together with the calculated ones for $r = 0.41$(solid areas). (c) $r$ (left axis) and $v$ in the bar region (right axis) as a function of $B$. (d) $C_{ch}$ and $C_x$ calculated using Eqs. (1) and (2) as a function of $B$. Inset shows a schematic of coupled EC circuit model.

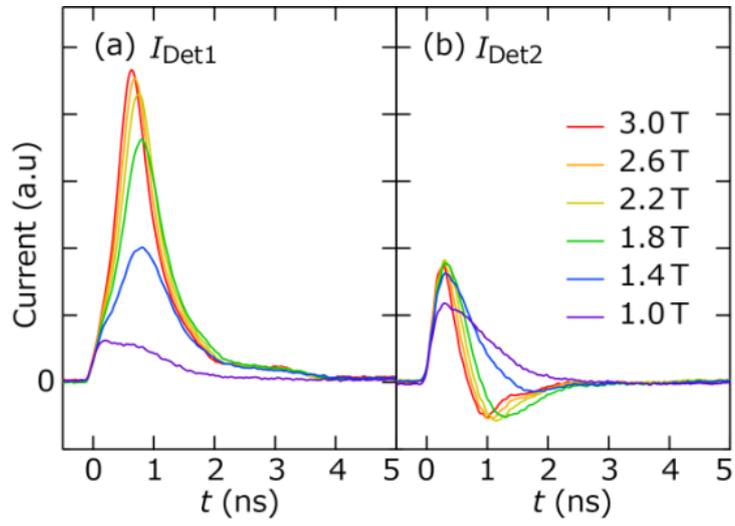

FIG. 4. (a) and (b) $I_{Det1}$ and $I_{Det2}$ as a function of $t$, respectively, for several values of $B$ around the transition region between the QH and non-QH states.

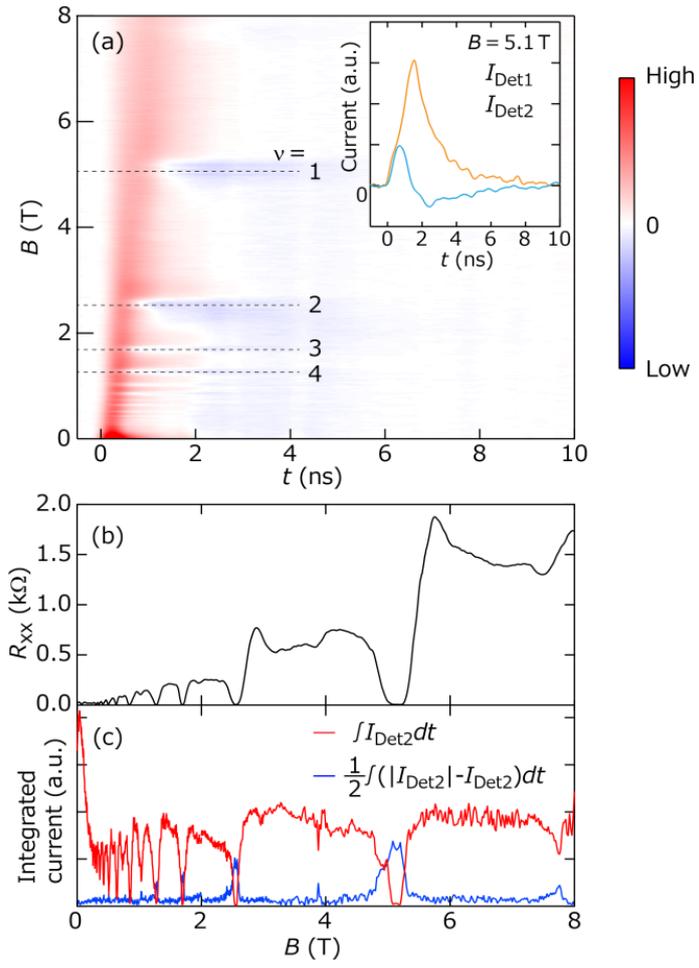

FIG. 5. (a) $I_{Det2}$ as a function of $t$ and $B$. Inset of shows $I_{Det1}$ and $I_{Det2}$ at $\nu = 1$ ($B = 5.1$ T). (b) and (c) $R_{xx}$ and the integration of $I_{Det2}$ as a function of $B$, respectively.

# Supplementary Information

**Calculation of current traces in the QHE regime**

In Fig. 3 of the main manuscript, we plotted $I_{\text{Det1}}$ and $I_{\text{Det2}}$ together with the calculated current traces. Here, we describe the details of the calculation, focusing on the evolution of the current waveform in the bar region. The dispersion of a coupled EMP wavepacket inside the bar region is

$$\omega(k) = \frac{1-r^2}{1+r^2}\left[\frac{\sigma_{xy}}{2\pi\epsilon^*\epsilon_0}\left(\ln\frac{2}{kw}+1\right)+v_{\text{D}}\right]k, \qquad (S1)$$

where $v_D$ is the drift velocity and $w$ is the transverse width of EMPs [S1,S2]. Then the current waveform at a position $x$ and time $t$ is given by,

$$\phi(x,t) = \int_0^\infty A(k)\exp i[kx-\omega(k)t]\exp\left(-\frac{t}{\tau}\right)dk, \qquad (S2)$$

where $A(k)$ is the Fourier transformation of the current waveform at the entrance of the bar region, and $\tau$ is the decay time. Multiple reflections and transmissions of the EMP wavepacket at both ends of the bar region yield a series of current pulses with charge $(1-r^2)Q, r^2(1-r^2)Q,\ldots$ on $I_{\text{Det1}}$ and $rQ, -r(1-r^2)Q,\ldots$ on $I_{\text{Det2}}$ at a constant time interval. Therefore, $I_{\text{Det1}}$ and $I_{\text{Det2}}$ become

$$I_{Det1} = (1-r^2)Q \times \phi(L,t) + r^2(1-r^2)Q \times \phi(3L,t) \qquad (S3)$$
$$+ \cdots$$

and

$$I_{Det2} = rQ \times \phi(0,t) - r(1-r^2)Q \times \phi(2L,t) - \cdots \qquad (S4)$$

respectively, where $L$ is the length of the bar region.

To obtain the current waveform in the main manuscript, we used $\epsilon^*$, $v_{\text{D}}$, and $w$ determined by previous work using graphene samples fabricated by similar processes [S3]. Using $r$, $Q$, and $\tau$ as tunable parameters, we reproduced $I_{\text{Det1}}$ and $I_{\text{Det2}}$. Note that the evolution of the EMP wavepacket outside of the bar region is also included in the calculation. We assumed that the initial wavepacket is a Gaussian with a width of 200 μm.

[S1] V. A. Volkov and S. A. Mikhailov, Zh. Eksp. Teor. Fiz. **94**, 217 (1988).
[S2] E. Berg, Y. Oreg, E.-A. Kim, and F. von Oppen, Phys. Rev. Lett. **102**, 236402 (2009).
[S3] N. Kumada, P. Roulleau, B. Roche, M. Hashisaka, H. Hibino, I. Petkovic, and D. C. Glattli, Phys. Rev. Lett. **113**, 266601 (2014).